\documentclass[a4paper,showpacs]{revtex4}
\usepackage{graphicx}
\usepackage{amssymb}
\usepackage{amsfonts}
\usepackage{amsmath}
\usepackage{hyperref}

\newcommand{\Meson}{\mathcal{M}}

\begin{document}

\title{On the possibility of revealing the transition of a baryon pair state to a six-quark confinement state}

\author{V.I.\,Komarov}
\email[E-mail: ]{komarov@jinr.ru}
\affiliation{Dzhelepov Laboratory of Nuclear Problems, Joint Institute for Nuclear Research, RU-141980 Dubna, Russia}

\begin{abstract}
Proton-proton collisions are considered to find favourable conditions for searching for the transition of a baryon pair state to a hexa-quark confinement state $(3q)+(3q)\rightarrow(6q)_\mathrm{cnf}$.
It is admitted that central $pp$ collisions in a definite range of the initial energy can lead to creation of an intermediate compound system where the hexa-quark dibaryon can be formed.
Criteria for selection of central collision events and for manifestation of the quark-structure dibaryon production are proposed.
\end{abstract}


\maketitle

\label{sec:intro}
\section*{Introduction}

Quark confinement in the nonperturbative regime remains one of the most urgent and fundamental problems within the modern Standard Model of strong and electroweak interactions.
It stays to be a great challenge to physics from the times of the confinement formulation~\cite{Chodos:1974, Wilson:1974} up to the recent studies~\cite{Kondo:2015, Roberts:2016}.
In spite of the need in an adequate strict theory of the quark confinement and extensive efforts to develop such a theory there is no generally accepted understanding of the confinement.
This absence of the mathematically well-defined SU(3)$_c$ gauge theory~\cite{Jaffe:2006} leads the experts in the field to believe that solution of this one of the most fundamental problems in the modern physics is unlikely to be found through theoretical analysis alone: a constructive feedback between experiment and theory is required.

One of the ways for that is a search for and study of hadrons with a non-convenient quarkcontent: tetra-quarks, penta-quarks and hexa-quarks.
The first two have become a hot point of modern hadronic physics in the last decade (see e.g.~\cite{Chen:2016}).
But a search for the latter has already a long history.
Two-baryon states as members of a multiplet in the aspect of the SU(3) symmetry were first considered in~\cite{Oakes:1963}.
The deuteron was taken as the lowest member of the multiplet and the unbound resonant baryon-baryon states as the higher ones.
This idea was developed further~\cite{Dyson:1964} in the classification of two-baryon states via the SU(6) theory of strongly interacting particles.
The lowest members of the isospin $T$, angular momentum $J$ multiplet were again the deuteron $(T=0,J=1)$ and the deuteron singlet state or the s-wave diproton (1,0).
The next, higher states, were the s-wave $N\Delta(1232)$ resonance (1,2) and the s-wave $\Delta(1232)\Delta$(1232) resonance (0,3), so far unobserved that time.
However, the extensive search for the dibaryon states only began after the arising of the quark-bag model expectations~\cite{Aerts:1978, Mulders:1980}.
The main criterion used in the searches for identification of dibaryons was a small width of the candidate states ($\Gamma\lesssim100$~MeV) comparing with the width of the low-mass baryon resonances~\cite{Locher:1986, Bakker:1994}.
The only important positive result of numerous experiments was establishment of three resonance states $^1\!D_2$, $^3\!F_3$, $^3\!P_2$ (see~\cite{Arndt:1983, Arndt:2007} and refs.\ therein), and recently $^3\!P_0$~\cite{Komarov:2016} in the $pp$ interactions.
The position of the corresponding poles of the S matrix in the complex energy plane close to the $\Delta N$ branching line and the width comparable with that of the $\Delta$(1232) resonance led to a common interpretation of the resonances as the conventional hadron states in the $\Delta N$ channel but not the quark dibaryons.
The attention to the quark-structure dibaryons was revived only several years ago with observation of the isoscalar $NN$ resonance in the energy region of the $\Delta(1232)\Delta(1232)$ excitation~\cite{Bashkanov:2009, Adlarson:2014}.
A relatively small width of the resonance, $\Gamma\approx70$~MeV, stimulated the authors to interpret the resonance as evidence of the genuine dibaryon.
However, the calculations~\cite{Gal:2013, Gal:2014} in a $\pi N\Delta$ model with three-body techniques applied to the Faddeev equations, resulted in the mass and width of the $T(J^P)=0(3^+)$ resonance that agreed well with the experimental values~\cite{Bashkanov:2009, Adlarson:2014}.
In addition, the parameter $r_0$ quantifying the spatial extension of the $\Delta N$ form factor, was found at a level of 1~fm.
This did not require the introduction of any short-range degrees of freedom, in particular quark-gluon ones.
A proper description of the $\Delta N$ resonances was also obtained in a similar model dealing with the meson-baryon approach~\cite{Gal:2014}.
Therefore, the nucleon-delta and delta-delta resonances observed up to now do not look like any short-range hexa-quark objects.
Meanwhile, development of chiral constituent quark models in the last two decades provided an essential advance in the understanding of the main features of the two-baryon systems involving the short-range interaction (see~\cite{Valcarce:2005, Ping:2009, Woosung:2015} and refs.\ therein).
Various models in this approach show a possibility of the transition
\begin{equation}\label{eq1}
(3q)+(3q)\rightarrow(6q)_\mathrm{cnf},
\end{equation}
where $(6q)_\mathrm{cnf}$ is a six-quark system confined in a hadron-like state.
Such states may be bound more deeply than the $N\Delta$ or $\Delta\Delta$ threshold states close to them and are therefore stable relative to a fast decay to these states.
This leads to the expected resonance behavior of the $NN$ interaction determined by the quark space, isospin and color degrees of freedom in the $(6q)_\mathrm{cnf}$ dibaryon.
The models are still unable to predict the exact values of the resonance masses and widths but do not exclude the width values less than 100~MeV.

The failure to observe such genuine dibaryon resonances might have two evident causes: first, the phenomenon is really absent in Nature, and second, the searches performed have not satisfied the proper conditions necessary for success.

\label{sec:The conditions favorable for the quark-structure dibaryon manifestation }
\section*{The conditions favorable for the quark-structure dibaryon manifestation }

What may be definitely affirmed concerning a lack of the necessary conditions in the known experiments is the following.
The experiments disregarded an obvious request for success of transition~\eqref{eq1}, the request of the \textbf{space overlapping} of both three-quark wave functions to create the six-quark system.

\begin{subequations}\label{eq2}
Such overlap cannot be directly reached in any peripheral collision of two nucleons.
It is a \textbf{central $NN$ collision} that is a correct way to get the overlap.
Therefore, the kinematical conditions for the successive experiment should satisfy criteria for the central $NN$ collision.
That means the impact parameter $R_\mathrm{imp}$ of the collision should be less than the size of the quark core of the nucleon.
This size is approximately evaluated as 0.4~fm:
\begin{equation}\label{eq2a}
R_\mathrm{imp}<0.4~\mathrm{fm}.
\end{equation}
This requirement can be realized if the transverse momentum transfer from the initial to the final nucleon is sufficiently high
\begin{equation}\label{eq2b}
q_{\perp}>h/R_\mathrm{imp}\approx500~\mathrm{MeV}/c.
\end{equation}

Strictly speaking, this requirement constrains only one projection of the impact parameter, namely, the projection onto the plane defined by the initial momentum $p_0$ and the momentum $p_\perp$ of the detected final state nucleon.
However, a short-range character of the forces providing the high-momentum transfer requires a small impact parameter also in the orthogonal plane.

The other way to constrain both projections of the impact parameter is to restrict the size \textbf{$r_\mathrm{int}$} of the $pp$ interaction volume.
It can be reached if the interaction is inelastic and generates secondaries with a sufficiently high mass $M$.
Then a relevant time interval $\Delta t$ is restricted by the relation $\Delta t \approx h/M$ and correspondingly $r_\mathrm{int}$ is restricted by the relation \textbf{$r_\mathrm{int}$}$\approx hc/M$.
Following~\eqref{eq2a}, one obtains
\begin{equation}\label{eq2c}
M \gtrsim 400~\mathrm{MeV}/c^2.
\end{equation}
\end{subequations}

The requirement of centrality is not yet sufficient for the overlapping: the evident obstacle is a short-range repulsion (SRR) in the nucleon-nucleon interaction.
This feature is of fundamental importance.
In general, it saves the nucleons of the ground state nuclear matter from sticking together after internuclear collisions and therefore prevents coalescence of nuclear matter to dense quark matter.
But at the same time, it forbids transition~\eqref{eq1} in the collisions where the c.m.s.\ energy is deficient to overcome the SRR.
The repulsion does not create an absolutely hard core at distances less than $\approx$ 0.4~fm.
According to modern constituent quark models, the repulsive interaction has a finite size potential $V(R)$ depending on a distance $R$ between the centers of the colliding nucleons.
So, the core is impermeable only at relatively low energies.
If the c.m.s.\ kinetic energy of the colliding nucleons is higher than $V(0)$, the nucleons are mutually penetrable and their content in a central collision can become joint at R = 0.
The model calculations give considerable scatter of the $V(0)$ estimates at a level of 0.5--1.0~GeV~\cite{Valcarce:2005, Bartz:1999, Ishii:2007}, which corresponds to variation of $\sqrt{s^\mathrm{min}}$ needed for overcoming the SRR at values of 2.4--2.9~GeV.
In the fixed-target experiments, the laboratory beam energy corresponds to $T_\mathrm{lab}^\mathrm{min}\approx$~1.1--2.5~GeV.
It means in particular that the resonances observed in the $pp$ collision at $0.7$~GeV~\cite{Komarov:2016} and in the $pn$ collision at $1.14$~GeV~\cite{Bashkanov:2009, Adlarson:2014} are most likely to be of a meson-baryon nature, the more so that criteria~\eqref{eq2} of the central collisions are not satisfied there.

At energies slightly higher than $T_\mathrm{lab}^\mathrm{min}$ the initial momentum of the participating constituent quarks, in average $P_\mathrm{cms}/3$, becomes totally spent for the mutual braking of the nucleons at the distance $R=0$.
The arising six-quark system is in the intermediate state $(6q)^*$ with about doubled baryon density and excitation energy $\sqrt{s}-2m_{N}$.
The baryon density is $\rho_{NN}=2/(4/3\pi r_c^3) \approx 4.3~\mathrm{fm}^{-3}$, which is higher than the critical baryon density providing the baryon deconfinement in the nuclear matter $\rho_c=0.85~\mathrm{fm}^{-3}$.
The energy density exceeds $\varepsilon_{NN} = 2m_N/(4/3\pi r_c^3)$ $\approx 4~\mathrm{GeV}~\mathrm{fm^{-3}}$, which is also higher than the critical energy density $\varepsilon_c=1$~GeV~fm$^{-3}$ necessary for the baryon deconfinement in nuclear matter.
So, the state of hadronic matter in the intermediate compound system $(6q)^*$ is a definitely deconfined quark-gluon state.
The system is unstable: it expands in space and loses its energy via the meson cooling.
If at any appropriate density and excitation it gets the confinement structure $(6q)_\mathrm{cnf}$ of a hadronic type, the process may acquire resonant behavior, and the system lives for a time $1/\Gamma$, where $\Gamma$ is the width of the resonance.
It is right a resonance that is the quark-structure dibaryon, a goal of the many-year search.
Such a scenario does not exclude a case where the transition $(6q)^*\rightarrow(6q)_\mathrm{cnf}$ may proceed immediately after the fusion of the nucleons without any additional energy loss.
Unfortunately, the above-mentioned chiral constituent quark models indicate only possible existence of the $(6q)_\mathrm{cnf}$ states with definite energy but do not try to describe the relevant dynamics of their construction.

With increase of the initial energy, the quarks of the intermediate system $(6q)^*$ conserve part of the initial momenta collinear to the collision axis.
If the quarks acquire the momentum higher than about 0.84~GeV/$c$ determining the proton formfactor, they leave the common 6-quark space and hadronize up to the leading nucleons in the fragmentation region of the $pp$ interaction.
A stage of the short-lived intermediate six-quark system is excluded there, and the formation of any $(6q)_\mathrm{cnf}$ state is also excluded.
Crude estimation of the corresponding energy gives the higher boundary $\sqrt{s^\mathrm{max}}\approx$~5.9--6.1~GeV and $T_\mathrm{lab}^\mathrm{max}\approx$~16.5--19.8~GeV.

Thus, the energy interval where the $2(3q)_\mathrm{cnf}\rightarrow(6q)_\mathrm{cnf}$ transition via the central $NN$ collision may be expected is rather limited
\begin{equation}\label{eq3}
\begin{array}{r@{~}c@{~}l}
\approx 2.5~\mathrm{GeV} < & \sqrt{s}       & \lesssim 6~\mathrm{GeV};\\
\approx 2~\mathrm{GeV}   < & T_\mathrm{lab} & \lesssim 20~\mathrm{GeV}.
\end{array}
\end{equation}

It is reasonable to choose the desirable final state of the nucleon pair in the form of a bound $np$ pair, a deuteron, or a quasi-bound $^1\!S_0$-state of the proton pair, ${pp}_s$.
(Further, for brevity, we only mention the deuteron.)
This choice is consistent with the lowest ground state of the Dyson-Xuong dibaryon multiplet~\cite{Dyson:1964} mentioned earlier.
More convincing justification of such a choice is the inelasticity of the whole process automatically following there: the bound nucleon pair has to be kinematically accompanied by a system $X$ of the produced particles: $p+p\rightarrow d+X$.
It excludes the perfectly studied elastic scattering $pp\rightarrow pp$ at large angles where no evidence for the dibaryon formation has been ever seen.
There is a drastic difference between the central elastic collision and the inelastic collision with the formation of a deuteron.
In the elastic scattering there is no immediate limitation on the impact parameter in the plane orthogonal to the scattering plane, and the process proceeds predominantly via a single scattering of the quarks only resulting in a change in the direction of the proton momentum.
It conserves the inherent quark structure of the incident protons and excludes the formation of the six-quark dibaryon.
In the case of the deuteron formation, the large invariant mass of the incident nucleon pair diminishes to a low value of the deuteron mass.
It requires a significant change in the relative momenta of the participating quarks, which leads to significant reconstruction of the total quark structure, necessary to form the final nucleon pair with a small relative momentum.
In addition, the elastic scattering excludes the use of the centrality criterion~\eqref{eq2c} determining the small radius of the interaction volume.
To get a maximum transversal momentum at a given energy the deuteron production angle of $90^{\circ}$ should be chosen.

Thus, the favorable conditions for the search for process~\eqref{eq1} can be obtained in a simple process
\begin{equation}\label{eq4}
p+p\rightarrow d|_{90^{\circ}}+X
\end{equation}
studied in energy region~\eqref{eq3}.

In addition to the high efficiency of process~\eqref{eq4} for manifestation of the transition of interest, which is due to the overlapping of incident protons, the process also has an other advantage.
It is a minimal contribution of the background peripheral processes.
Indeed, the latter proceed with dominant emission of final state nucleons with opposite directions of their c.m.s.\ momenta, which results in their predominantly high relative momentum.
Emission of secondaries mainly at small angles relative to the collision axis minimizes their yield near the angle of $90^{\circ}$ too.

In~\eqref{eq4}, $X$ denotes a system $\Meson$ of mesons, since production of baryon-antibaryon pairs in energy region~\eqref{eq3} is forbidden or strongly suppressed.
These final mesons are for the most part pions from direct generation or from decay of intermediate mesons.
A small admixture of kaons also exists in $\Meson$.

Detection of only a deuteron (or a $^1\!S_0$ diproton) in reaction~\eqref{eq4} includes study of different channels of the $6q$ confined decay:
\begin{subequations}\label{eq5}
\begin{align}
(6q)_\mathrm{cnf}&\rightarrow d+\pi; \label{eq5a}\\
(6q)_\mathrm{cnf}&\rightarrow d+2\pi; \label{eq5b}\\
(6q)_\mathrm{cnf}&\rightarrow d+\rho; \label{eq5c}\\
&\cdots \nonumber \\
(6q)_\mathrm{cnf}&\rightarrow d+n_\mathrm{max}\pi. \tag{5n}\label{eq5n}
\end{align}
\end{subequations}
Certainly, a single heavy meson can be produced in~\eqref{eq5n} instead of the $n_\mathrm{max} \pi$.

A signature for the transition of interest may be a resonance peak (or peaks) in the energy dependence of the inclusive differential cross section of the deuteron (or $^1\!S_0$ $pp$) emission in reaction~\eqref{eq2}.
The cross section may be chosen to include all channels~\eqref{eq5}, single channel~\eqref{eq5a} or a definite region of the selected meson system invariant masses.
Such variation may give additional information on the dynamics of decay~\eqref{eq5}.
Of special interest is extreme channel~\eqref{eq5n}, where all the kinetic energy of the colliding protons is spent to form the mesonic field.

\begin{figure}[ht]\centering
\includegraphics[width=127mm]{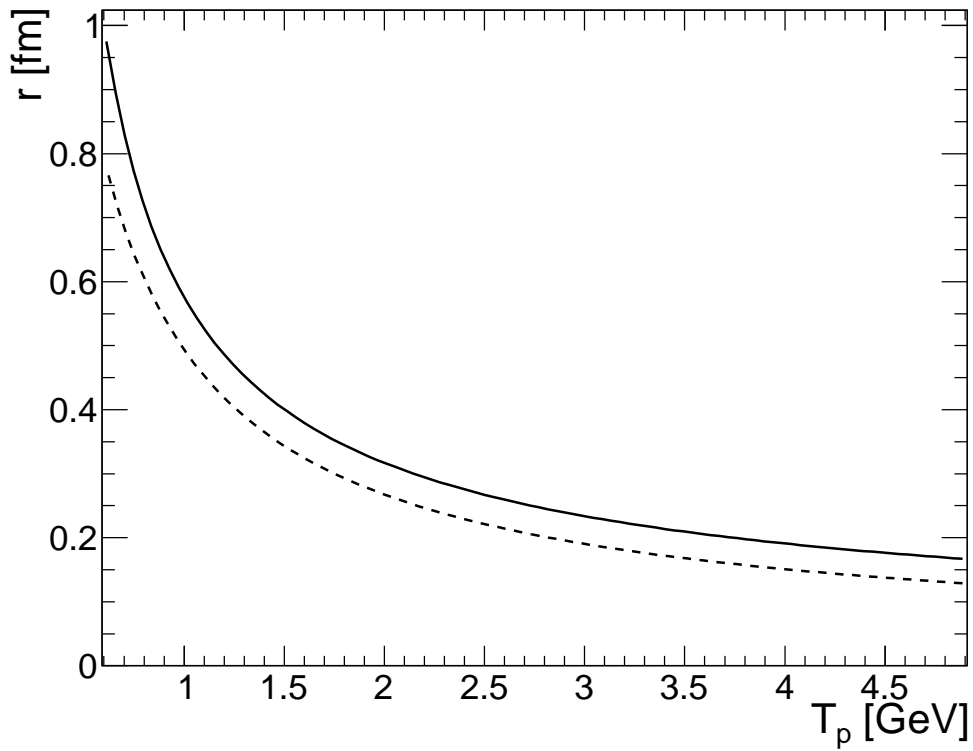}
\caption{\label{centralities}Centrality in the process $pp\rightarrow dX$. Lines are the impact parameter $r$ for the channel $pp\rightarrow d|_{90^{\circ}} + \pi^+$ (solid) and the $pp\rightarrow dM_\mathrm{max}$ (dashed).}
\end{figure}

Reaction~\eqref{eq4} allows centrality criterion~\eqref{eq2} to be fulfilled in the energy region of interest~\eqref{eq3}.
Figure~\ref{centralities} shows dependence of $r_{\perp}$ on the initial energy $T_p$ for the channel $pp\rightarrow d\pi^+$ producing the highest deuteron momenta.
It is seen that $r_{\perp}$ is smaller than 0.5~fm at all energies higher than 1.5~GeV.
As the invariant mass of the meson system $\Meson$ increases, the deuteron momentum diminishes but the centrality criterion does conserve since the meson production volume simultaneously decreases.
The radius of this volume becomes smaller than 0.5~fm at the energy $T_p>1$~GeV for the kinematical extremity of the maximal mass production, $M_\mathrm{max}=\sqrt{s}-2m_p$.
It is also seen from Figure~\ref{centralities}.
The criterion of the interaction volume smallness can be expressed in terms of the invariant momentum transfer too.

The proposed way for observation of the $6q$ dybaryon confinement states also allows searching for    resonances in a low mass region being beyond the immediate reach because of the SRR counteraction.
Such states can be observed via a sequence of processes
\begin{equation}
p+p\rightarrow (6q)^* \rightarrow (6q)_\mathrm{cnf} + \pi \rightarrow (d+p)+\pi, \label{eq6}
\end{equation}
where the first step proceeds at the energy sufficient to overcome SRR.
The expected resonance can manifest itself as a peak in the deuteron energy spectra at a fixed incident energy.

It is worth noting here that the criterion of the short-range interaction volume considered above may have a more general use than in the case of the quark-structure dibaryon formation.
Indeed, this criterion may be a strong argument for identification of other ``elementary'' hadrons with a nonstandard quark content.
A problem of distinguishing between genuine elementary hadron and a composite state of two normal hadrons was recognized a long time ago~\cite{Vaughn:1961, Weinberg:1965}.
In~\cite{Weinberg:1965} S.~Weinberg formulated the evidence that the deuteron is not an elementary particle.
The way suggested by him was developed later~\cite{Baru:2004, Kamiya:2017} for the case of quasibound unstable particles.
At present, this problem to distinguish composite from elementary particles becomes urgent in connection with observation of tetraquark and pentaquark candidates~\cite{Chen:2016}.
Unfortunately, Weinberg's criterion requires knowledge of the low-energy scattering parameters of the composed hadrons, which is usually scarce or absent.
In this situation, the candidates for ``elementary particle``, genuine quark-structure particle state should be kinematically tested right for the short-range production volume.

\label{sec:Existing and desired experimental data}
\section*{Existing and desired experimental data}

\begin{figure}[ht]\centering
\includegraphics[width=127mm]{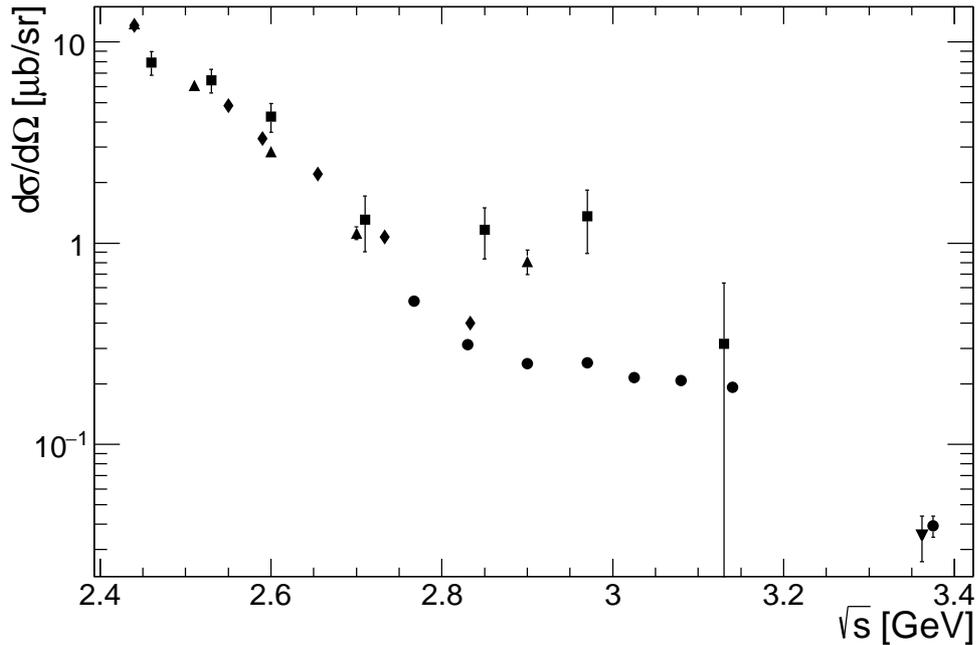}
\caption{\label{dsg90log}Data on the differential cross section of the reaction $pp\rightarrow d + \pi^+$ at $90^{\circ}$ in the range $T_p=$~2.4--3.4~GeV.
Circles $\bullet$ denote the results from~\cite{Anderson:1974}, squares $\blacksquare$ correspond to the results from~\cite{Dekkers:1964}, triangles $\blacktriangle$ correspond to the results from~\cite{Heinz:1968}, down triangle $\blacktriangledown$ corresponds to the result from~\cite{Ruddick:1968}, and diamonds $\blacklozenge$ correspond to the results from~\cite{Yonnet:1993}.}
\end{figure}

Despite a vast amount of experiments devoted to the deuteron production in the $NN$ collisions at intermediate and high energies, no systematic measurements of process~\eqref{eq4} in energy region~\eqref{eq3} are known.
Even the most explored reaction $pp\rightarrow d\pi^+$ presents rather scarce and contradictory data compiled in Figure~\ref{dsg90log}.
The most perfect of them, the old experiment~\cite{Anderson:1974}, was performed in a full angular interval 0$^{\circ}$--90$^{\circ}$ at the beam energies $T_p =$~1.45--4.15~GeV.
The energy dependence of the differential cross section at $90^{\circ}$ does not exclude presence of a wide bump around $\sqrt{s} \approx 3.0$~GeV; however, its manifestation needs at least expansion of the data in the energy region $T_p =$~3.5--6.0~GeV.
There is a set of measurements of the reaction $p + p\rightarrow d + X$ cross sections at energies of 3--23~GeV, but all of them are at rather small angles (see~\cite{Amaldi:1972} and refs.\ therein).
The energy dependence of the $pp\rightarrow dX$ cross section at the c.m.s.\ angles close to $90^{\circ}$ cannot be obtained from the existing data.

The first step of the experiments on the problem considered may be inclusive measurements of the reaction~\eqref{eq4} cross section.
It is worth stressing that the relevant measurements are relatively simple and do not require much of expenditures.
Indeed, it supposes an experimental device of a standard kind where deuterons emitted from proton-proton collisions traverse a magnetic field region for the momentum analysis, two planes of counters for a time-of-flight measurement, and planes of multiwires proportional chambers.
The deuterons are stopped in the $\Delta E-E$ counters.
Since the deuteron energies are in a range from about tens of MeV up to several hundreds of MeV, the needed sizes of the magnetic field region and of the detectors are quite modest.
The present accelerators definitely provide luminosities for such measurements with the cross sections in the 1--0.001~$\mu$b/sr range both in the collider and the fixed-target mode.
If the resonances considered are observed the measurements may be extended to a wider angular acceptance range, study of the correlations of secondaries and the polarization observables.

Disappointing possible absence of the resonances has nevertheless its own significance.
First, it challenges the chiral compound quark models expecting the six-quark confinement states and requires elucidation of reasons forbidding such states.
Second, the experiments open a study of central nucleon-nucleon collisions at energies providing production of highly excited intermediate quark-gluon compound states.
Such collisions are the elementary process of the two-baryon deconfinement with subsequent reconstruction of baryons.
Features of the processes are terra incognita at present.
The study may give immediate experimental information to promote advance of the desired nonperturbative confinement theory.
It is worth stressing that nucleon-nucleon collisions studied for a long time in a vast number experiments under a great variety of conditions nevertheless have not yet been thoroughly studied in the conditions discussed above.

\label{sec:Summary}
\section*{Summary}

Symmetry considerations of two-baryon systems and calculations in different models of their QCD structure indicate a possibility of the transition between two incident nucleons and a hadron with the baryon number B=2 with a hexa-quark confinement structure.
There is no commonly accepted experimental evidence of such transitions.
It makes a significant challenge of fundamental importance to perform experiments with the aim to reveal the transitions.
The experiments should be done in conditions favorable for this task.
A study of the reaction $p + p\rightarrow d|_{90^{\circ}} + \Meson$ looks promising for that.
The study may give currently unavailable experimental information to promote advance of the desired nonperturbative confinement theory.

It should be noted in conclusion that studies with a similar motivation may be performed with the processes $p + n \rightarrow d|_{90^{\circ}} + \Meson$ and $e + d \rightarrow d|_{90^{\circ}} + \Meson + e'$.
The systems more complicated than dibaryons can be also studied in central collisions via the processes as $p + d \rightarrow{}^3\mathrm{He}|_{90^{\circ}} + \Meson$.

\begin{acknowledgments}
The author is thankful to V.A.~Bednyakov, V.V.~Burov, N.I.~Kochelev, A.V.~Kulikov, and D.A.~Tsirkov for the interest expressed in the stimulating discussions. I am also sincerely indebted to B.~Baimurzinova, A.~Kunsafina, and Zh.~Kurmanaliev for their help in preparing the publication.
\end{acknowledgments}

\bibliographystyle{aipnum4-1}
\bibliography{Komarov_biblio_arxiv}

\end{document}